\documentclass[12pt]{article}
\usepackage{amsmath,amssymb,mathrsfs,bm}
\usepackage{subfigure}
\usepackage{graphicx}
\begin{document}

\begin{center}
\begin{large}
{\bf Chaotic Information Processing by Extremal Black Holes}
\end{large}

\vskip1.0truecm

Minos Axenides$^{(a)}$\footnote{E-Mail: axenides@inp.demokritos.gr}, Emmanuel  Floratos$^{(a,b)}$\footnote{E-Mail: mflorato@phys.uoa.gr} and Stam Nicolis$^{(c)}$\footnote{E-Mail: stam.nicolis@lmpt.univ-tours.fr}

\vskip0.3truecm

{\sl 
{ }$^{(a)}$ Institute of Nuclear and Particle Physics, NCSR ``Demokritos''\\
GR--15310, Athens, Greece\\ \vskip0.1truecm
{ }$^{(b)}$ Department of Physics, University of Athens\\
GR--15771, Athens, Greece\\ \vskip0.1truecm
{ }${(c)}$ CNRS--Laboratoire de Math\'ematiques et Physique Th\'eorique (UMR7350)\\
F\'ed\'eration de Recherche ``Denis Poisson'' (FR2964)\\
D\'epartement de Physique, Université de Tours\\
Parc Grandmont, 37200 Tours, France
} 
%\date{VIIth Black Hole Workshop, Aveiro, PT 19 December 2014}

%\thanks{Work based on arXiv:1306.5670 and in progress,with M. Axenides  and E. G. Floratos.}

%\maketitle

\end{center}
\begin{abstract}
We review an explicit regularization of the AdS$_2$/CFT$_1$ correspondence, that preserves all isometries of bulk and boundary degrees of freedom. This scheme is useful to characterize  the space of the unitary evolution operators that describe the dynamics of the microstates of extremal black holes in four spacetime dimensions. Using techniques from algebraic number theory to evaluate the transition amplitudes, we remark that the regularization scheme expresses the fast quantum computation capability of black holes as well as its chaotic nature.    
\end{abstract}
\section{Introduction}\label{intro}
Black holes are classical objects,  since they are solutions of Einstein equations--however, they provide the opportunity to probe 
quantum aspects of spacetime, too. In particular, while they possess infinite entropy, if probed by classical observers, it turns out that they possess 
finite entropy, if probed by quantum observers. This statement implies that, while the classical space of states is infinite dimensional, the quantum space of states is finite dimensional. However it was also found that, for generic values of mass and charges, black holes possess, also, temperature and, consequently, emit radiation. This statement implies that black hole entropy doesn't remain constant, but is transferred from the black hole to the radiation. This process is understood, still, only semi-classically and isn't under full quantitative control for arbitrarily long times~cf. \cite{barbon_rabinovici} for a review--in particular the issue of black hole evaporation can't be reliably addressed, yet.

However there are dynamical issues, that involve quantum properties of black holes, that can be addressed quantitatively and lead to interesting insights for and from  information theory. These pertain to the description of the dynamics of the black hole microstates, that provide the statistical mechanical interpretation of the entropy of the black hole, when mass and charges of the latter satisfy the extremality condition. In that case the temperature of the black hole vanishes, it does not radiate and its entropy is fixed--and non-zero. To describe the quantum dynamics of the black hole microstates then amounts to describing  the space of unitary operators that map the (finite--dimensional) space of states to itself and to computing the transition amplitudes. What hasn't been widely appreciated is that these problems admit an explicit solution, that involves insights from algebraic number theory and quantum computing. Furthermore, this framework allows us to address the question of how fast information is processed and provide evidence for the ``fast scrambling conjecture''. Finally, the framework we shall review in the following, allows us to, at least, envisage a way of addressing the case where the size of the space of states can vary, in a controlled way and, thus, perhaps, probe issues beyond the extremal limit. 

The plan of the paper  is as follows:

In section~\ref{BHgeom} we recall that extremal black holes have a near horizon geometry described as a space of the form AdS$_2\times K$, where $K$ is the space of charges. To be specific, we shall use the extremal Reissner--Nordstr\"om black hole as an example.

The AdS$_2$ describes, on the one hand, the geometry of the phase space of the black hole microstates and, on the other hand the radial and temporal dimensions of the spacetime geometry. Indeed there is a 1--to--1 correspondence between the points of the former and those of the latter. This property is very particular to extremal black holes--it isn't, of course, true for an arbitrary quantum system. 

 This implies that a classical probe ``sees'' a continuum geometry and probes an infinite number of states, while a quantum probe, that probes a finite number of states, also ``sees'' a discrete, finite,  geometry. However the way the quantum probe does this is consistent with the isometries of AdS$_2$. 

This is realized in the string theoretic construction, by a regularization of the AdS$_2$ geometry. In our construction, this is realized through, another,   ``modular'' regularization, that was the subject of a recent publication and whose salient points we briefly discuss. 

This construction allows us to write the most general, unitary operators, that map the discrete space of states to itself and thus can  represent the one time step evolution operators of the probes.

In section~\ref{cats} we show that these operators represent the consistent quantization of Arnol'd cat maps and this allows us to apprehend that the quantum information processing carried out by the extremal black hole is chaotic, while being, also, unitary.  Using tools from algebraic number theory, we show that it is possible to express the transition amplitudes in terms of Gauss sums and to study how an initial overlap among a small number of states, very rapidly, spreads to all states. 

In section~\ref{conclusions}  we discuss some ideas for directions of further inquiry.

\section{Extremal black hole geometry as a spacetime}\label{BHgeom}
The example we shall use is the Reissner--Nordstr\"om black hole in $D=4$ spacetime dimensions. Its metric is given by the expression 
\begin{equation}
\label{RNmetric}
ds^2 = -h(r) dt^2 + \frac{dr^2}{h(r)} + r^2d\Omega^2
\end{equation}
where $d\Omega^2$ is the metric of $S^2$ and 
\begin{equation}
\label{metricfunction}
h(r) = \frac{(r-r_-)(r-r_+)}{r^2}
\end{equation}
is the metric function and the radii, $r_\pm$ of the event horizons are given, in units where $G=1, c=1$, by 
\begin{equation}
\label{reventhor}
r_\pm\equiv M\pm\sqrt{M^2-Q^2}
\end{equation}
where $M$ is the mass and $Q$ the total charge (in general, electric and magnetic) of the black hole. 

The near horizon limit is defined through the relations~\cite{Sen12} (using slightly different conventions) 
\begin{equation}
\label{nhern}
\begin{array}{ccccc}
\displaystyle 2\lambda\equiv r_+ - r_- & & \displaystyle {\sf t}\equiv\frac{\lambda t}{r_+^2} & & 
\displaystyle {\sf r}\equiv\frac{2r-r_+ -r_-}{2\lambda}
\end{array}
\end{equation}
namely, when $\lambda\to 0$, keeping ${\sf r}$ and ${\sf t}$ fixed. These parametrize the radial and temporal directions in this limit. 

The metric, in the ${\sf r}$ and ${\sf t}$ variables, takes the form
\begin{equation}
\label{metricnehg}
ds^2= \underbrace{r_+^2\left(-({\sf r}^2-1)d{\sf t}^2 + \frac{d{\sf r}^2}{{\sf r}^2-1}\right)}_{\mathrm{AdS}_2} + \underbrace{r_+^2\left(d\theta^2 +\sin^2\theta d\phi^2\right)}_{S^2}
\end{equation}
known as the Bertotti--Robinson metric: the radial and time directions describe an AdS$_2$ space and the angular directions an $S^2$ space. The radii of the two are both equal to $r_+=r_-$.

In this, extremal, limit, the black hole doesn't radiate, since the Hawking temperature vanishes. Therefore it is possible to study the dynamics of the black hole microstates, that contribute to its entropy. The calculation involves the partition function of the fields, that give rise to such states in the black hole geometry. In string theory  the partition function is computed by summing over  the fields  in the full near--horizon geometry, that is of the form AdS$_2\times K$, where $K$ is a compact space that includes an $S^2$ factor. In the AdS$_2$/CFT$_1$ correspondence, this partition function is calculated by the partition function of the conformal field theory on the boundary--which, in this case, is, classically, disconnected, while quantum effects lead to entanglement. And, of course, boundary conditions must be imposed. What is relevant for the present contribution is that AdS$_2$ is a noncompact manifold, therefore a regularization must be introduced, to render the calculations well--defined. In ref.~\cite{Sen12,dabholkaretal} the consequences in one regularization scheme have been  studied. In ref.~\cite{AFN} another regularization scheme was presented. We shall review it in the following section and show how it highlights relations to quantum information processing, whose relevance for the description of the dynamics of the black hole microstates and their probes has recently become the topic of intense study~\cite{QuInfo15,Chaos2}.

\section{Extremal black hole geometry as a phase space}\label{cats}
The idea behind the regularization scheme, proposed in ref.~\cite{AFN}, is that the finite entropy of black holes implies that the number of states, accessible in a fixed charge sector of an extremal black hole, is finite. This space has the geometry of a one--sheeted hyperboloid, an AdS$_2$ manifold. To describe this phase space in a way that preserves its isometries, we  proposed a ``modular discretization'' of this manifold, i.e. to replace the equation 
\begin{equation}
\label{AdS2classical}
x_0^2 + x_1^2 - x_2^2=1
\end{equation}
over the real numbers, by the equation 
\begin{equation}
\label{AdS2quantum}
x_0^2 + x_1^2-x_2^2\equiv\,1\,\mathrm{mod}\,N
\end{equation}
where $N$ is number of microstates and $x_0,x_1,x_2$ now take values in the set $\mathbb{Z}/N\mathbb{Z}$. For $N$ a prime, $p$, this becomes the finite field, $\mathbb{F}_p$.

In this approach the bulk geometry is described by the ``finite geometry'', AdS$_2[p]=SL(2,\mathbb{F}_p)/SO(1,1,\mathbb{F}_p)$ and the conformal field  theory on the boundary by the ``finite geometry'' of the projective line. 

\begin{figure}[thp]
\begin{center}
\subfigure{\includegraphics[scale=0.5]{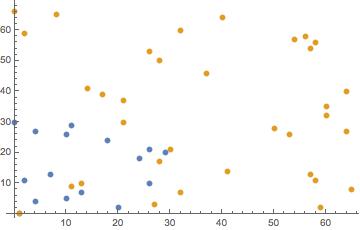}}
\subfigure{\includegraphics[scale=0.5]{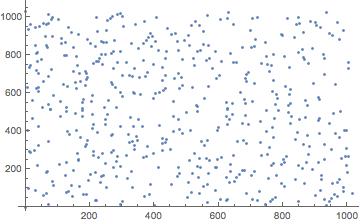}}
\end{center}
\caption[]{The time--like circle, $x_2\equiv 0\,\mathrm{mod}\,p$, for $p=31,67,1031$.}
\label{pmodcircle_fig}
\end{figure}

More precisely, the AdS$_2[p]$/CFT$_1[p]$ correspondence is realized as follows: 

A point, $(x_0,x_1,x_2)$ of the classical manifold (which does correspond to a matrix, $X\in SL(2,\mathbb{R})/SO(1,1)$), corresponds to a matrix, $X\in SL(2,\mathbb{F}_p)/SO(1,1,\mathbb{F}_p)$. The classical dynamics is described by the Weyl map, 
\begin{equation}
\label{weylmap}
X_{n+1}={\sf A}X_n{\sf A}^{-1}\mathrm{mod}\,p
\end{equation}
To get an idea of how ``chaotic'' such points look, compared to the smooth, classical, geometry, we display the time--like ``circle'', $x_2\equiv 0,\mathrm{mod}\,p$, thus the solutions to the equation $x_0^2+x_1^2\equiv 1\,\mathrm{mod}\, p$ in fig.~\ref{pmodcircle_fig}. These points certainly don't seem to belong on any circle--nevertheless they do encode its properties mod $p$, highlighting that the large $p$ limit is quite subtle.

In the quantum picture, the matrices $X$ label the states and the evolution operators become unitary  matrices, $U(A)$.  

In the AdS$_2[p]$ bulk, the stability group of the ground state  is the scaling subgroup, 
\begin{equation}
\label{scalingsubrgoup}
{\sf D}(\lambda)=\left\{\left.\left(\begin{array}{cc} \lambda & 0 \\ 0 & \lambda^{-1}\end{array}\right)\right|\lambda\in\mathbb{F}_p^\ast\right\}
\end{equation}
and the ground state is the common eigenvector, for all $\lambda\in\mathbb{F}_p^\ast$, of the unitary dilatation operator, 
\begin{equation}
\label{unirarydilatn}
U({\sf D}(\lambda))_{k,l}=(-\lambda|p)\left\{\begin{array}{c} 1 \\ -1\end{array}\right\}\delta_{k,\lambda^{-1}l}
\end{equation}
where $(-\lambda|p)$ is the Jacobi/Legendre symbol, the factor $\pm 1$ distinguishes  the case of a prime of the form $4q+1$
 from that of the $4q-1$ and $k,l\in\mathbb{F}_p$.

In components: 
\begin{equation}
\label{groundstatebulk}
\left|0\right\rangle_{\sf D}=\frac{1}{\sqrt{p}}\left(\underbrace{1,1,\cdots 1}_p\right)
\end{equation}
and the excitations,$\left\{|h\rangle\right\}$ are given, in components, by the expression
\begin{equation}
\label{bulkexcitations}
\left\langle k| h\right\rangle =\frac{1}{\sqrt{p}}\left( \left(a-b\mu\right)|p\right)\omega_p^{\frac{k^2}{2x_-}}
\end{equation}
where $k,\mu=0,1,2,3,\ldots,p-1$, $a^2+b^2\equiv\,1\,\mathrm{mod}\,p$ and $x_-\equiv (a-b\mu)(a\mu+b)^{-1} \,\mathrm{mod}\,p$. 

On the conformal boundary the stability group of the ground state is the Borel subgroup,
\begin{equation}
\label{boundary_stability}
\mathfrak{B}_p\equiv\left\{\left.  
\left(\begin{array}{cc} \lambda & b \\ 0 & \lambda^{-1}\end{array}\right)
\right|\lambda\in \mathbb{F}_p^\ast,b\in\mathbb{F}_p\right\}
\end{equation}
The ground state, $|0\rangle_B$, is given, in components,  by the expression
\begin{equation}
\label{boundary_gs}
\left\langle k|0\right\rangle_B=\delta_{k,0} 
\end{equation}
and the excitations, $\left\{|h\rangle_B\right\}$, by the expression
\begin{equation}
\label{boundary_excit}
\left\langle k| h\right\rangle_B=\frac{1}{\sqrt{p}}(-2b|p)\left\{\begin{array}{c}1\\ \mathrm{i}\end{array}\right\}\omega_p^{-\frac{x_+k^2}{2}}
\end{equation}
where $x_+\in\mathbb{F}_p^\ast$. 

Along with the ``points at infinity'', the number of states, in the bulk and on the boundary, is the same and operators can be explicitly constructed~\cite{AFN}.

The dynamics is, therefore, described by the action of the evolution operators on the states in the bulk and on the boundary. So any transition amplitude is given by the expression
\begin{equation}
\label{transition_amplitude}
\left\langle\psi_t|\chi_{t'}\right\rangle = \left\langle 0\right| \left[U({\sf A})^t\right]^\dagger{\mathcal O}U({\sf A})^{t'}\left|0\right\rangle
\end{equation} 
where $|\psi_t\rangle = U({\sf A})^t|0\rangle$. The calculation of this expression reduces to the evaluation of Gauss sums~\cite{AFN,AFN9598}. These computations are fast since the group representation property for the unitary evolution operators means that 
\begin{equation}
\label{groupfast}
\left[U({\sf A})\right]^t=U({\sf A}^t)
\end{equation}
Therefore, instead of multiplying $p\times p$ matrices $t$ times, one is multiplying $2\times 2$ matrices, $t$ times.

For any element, 
$$
{\sf A}=\left(\begin{array}{cc} a & b\\ c & d\end{array}\right)\in SL(2,\mathbb{F}_p)
$$
the unitary evolution operator, $U({\sf A})$, is given by the expression~\cite{AFN,AFN9598}
\begin{equation}
\label{UofA}
U({\sf A})_{k,l}=\frac{1}{\sqrt{p}}(-2c|p)\left\{\begin{array}{c} 1 \\ \mathrm{i}\end{array}\right\}\omega_p^{-\frac{ak^2-2kl+dl^2}{2c}}
\end{equation}
When $c\equiv 0\,\mathrm{mod}\,p$ the appropriate expression is known~\cite{AFN,AFN9598}. With these tools one can compute any transition amplitude in this regularization scheme.

\section{Conclusions}\label{conclusions}
We have reviewed a new regularization scheme for the partition function of the black hole microstates, for the case of extremal black holes and have constructed the state spaces on the boundary and the bulk finite geometries for any, prime, value.  
For any, odd, values of the exponential of the entropy, the construction of the states and evolution operators can be generalized  using the tools developed in refs.~\cite{AFN9598} and the relation with the tools presented in ref.~\cite{dabholkaretal} certainly deserves further study. Indeed, the asymptotic behavior, for large $p$, should correspond to the case of large charges. It should be stressed that this is a construction, {\em ab initio}, for bulk and boundary, that's valid for any value of $p$ and, thus, can describe quantum and semi--classical effects in the appropriate limits.

With the same tools it is, also, possible to describe observers, entangled with the black hole states. The unitary operates described here are, in fact, the gates for qubits and can be, also, studied using quantum information theory. There has been considerable work in this direction~\cite{QuInfo15,Chaos2} and we look forward to report on further developments in forthcoming work. 

{\bf Acknowledgements:} It's a great pleasure to thank the organizers of the VIIth Aveiro Workshop on Black Holes for a very stimulating meeting. SN would also like to thank G. W. Gibbons and E. Rabinovici  for sharing their insights on  black holes, through discussions and lectures in Tours and Paris.

\end{document}